\documentclass[aps,pre,twocolumn,groupedaddress,showpacs,floatfix]{revtex4}
\usepackage{dcolumn}
\usepackage{amsmath}
\usepackage{amssymb}
\usepackage{graphicx}
\usepackage{bm}
\usepackage[T1]{fontenc}
\usepackage{color}
\usepackage{url}
\usepackage[bf]{subfigure}
\usepackage{rotating}
\usepackage{algorithm}
\usepackage{algorithmic}

\begin{document}

\title{Synchronization in clustered random networks}

\author{Thomas Kau\^{e} Dal'Maso Peron}
\email{thomas.peron@usp.br}
\affiliation{Instituto de F\'{\i}sica
de S\~{a}o Carlos, Universidade de S\~{a}o Paulo, Av. Trabalhador
S\~{a}o Carlense 400, Caixa Postal 369, CEP 13560-970, S\~{a}o
Carlos, S\~ao Paulo, Brazil}
\author{Francisco A. Rodrigues}
\email{francisco@icmc.usp.br}
\affiliation{Departamento de Matem\'{a}tica Aplicada e Estat\'{i}stica, Instituto de Ci\^{e}ncias Matem\'{a}ticas e de Computa\c{c}\~{a}o,
Universidade de S\~{a}o Paulo, Caixa Postal 668,13560-970 S\~{a}o Carlos,  S\~ao Paulo, Brazil}
\author{J\"urgen Kurths}
\affiliation{Potsdam Institute for Climate Impact Research (PIK), 14473 Potsdam, Germany}
\affiliation{Department of Physics, Humboldt University, 12489 Berlin, Germany}
\affiliation{Institute for Complex Systems and Mathematical Biology, University of Aberdeen, Aberdeen AB24 3UE, United Kingdom}

\begin{abstract}
In this paper we study synchronization of random clustered networks consisting of Kuramoto oscillators. More specifically, by developing a mean-field analysis, we find that the presence of cycles of order three does not play an important role on network synchronization, showing that the synchronization of random clustered networks can be described by tree-based theories, even for high values of clustering. In order to support our findings, we provide numerical simulations considering clustered and non-clustered networks, which are in good agreement with our theoretical results. 
\end{abstract}


\maketitle

\section{Introduction}

Synchronization processes have attracted the interest of scientists for centuries and is in the focus of intense research today~\cite{Arenas08:PR}. This collective phenomena has been observed in biological, chemical, physical, and social systems~\cite{Pikovsky03, Acebron05:RMP}. Many works have verified that the dynamics of synchronization depends on the connectivity pattern of networks~\cite{Arenas08:PR}. For instance, when the natural frequency distributions are unimodal and even, the critical coupling depends on the ratio between the first and second statistical moments of the degree distribution~\cite{Ichinomiya04:PRE,lee2005synchronization,Restrepo05:PRE}. In addition, for networks in which there is a positive correlation between the network structure and dynamics, the critical coupling has an inverse dependence on the network average degre~\cite{Peron12}.

However most of the analytical results of networks consisting of Kuramoto oscillators have been obtained for uncorrelated networks generated through the configuration model~\cite{Ichinomiya04:PRE,lee2005synchronization,Restrepo05:PRE}, which generates networks with arbitrary degree distributions by randomly connecting the nodes according to a specified degree sequence. One of the main properties of this model is that in the thermodynamic limit, i.e., $N\rightarrow \infty$, the probability of occurrence of cycles of order three tends to zero. Such probability can be quantified through the clustering coefficient $C$, defined as
\begin{equation}
C = \frac{3 \times \left(\mbox{number of triangules in the network} \right)}{\mbox{number of connected triples}}=\frac{3N_{\triangle}}{N_3}.
\label{Eq:clustering_coeff}
\end{equation} 
In addition to the configuration model, the clustering coefficient also vanishes for Erdos-Renyi (ER) and Barabasi-Albert (BA) networks when $N\rightarrow \infty$. Therefore, most of the theoretical results concerning the Kuramoto model are derived for networks that have locally tree-like structures, i.e., $C\rightarrow 0$. However, most real-world networks have clustered topologies~\cite{Costa11:AP}. Thus, an analysis of the Kuramoto model on networks with clustering is necessary to model real-world synchronization with more accuracy.

The objective of the current work is to study the dynamics of synchronization on clustered random networks and perform a comparison with non-clustered ones. More specifically, we develop a mean-field theory for the configuration model proposed independently by Newman~\cite{newman2009random} and Miller~\cite{miller2009percolation}, which generates networks with $C>0$ even in the limit of large networks. Such analysis is compared with the mean-field theory developed for locally tree-like networks. Our results show that the mean-field theory for networks with low values of triangles can be applied with certain accuracy on clustered networks, indicating that the presence of cycles of order three does not influence the network synchronization.  This result is in agreement with previous works~\cite{melnik2011unreasonable,gleeson2010clustering,hackett2011cascades,accuracy2012gleeson}, which observed that the clustering coefficient does not play an important role in other dynamical process, such as bond-percolation, rumor and epidemic spreading, provided that the networks have low values of the average shortest path length. 

In Sec.II we briefly describe the configuration model proposed in~\cite{newman2009random} and~\cite{miller2009percolation}. In Sec.III we derive a sufficient condition for synchronization through mean-field approximation and in Sec.IV we compare numerical and theoretical results and give our conclusions.

\section{Random clustered networks}

In the standard configuration model, the network is generated through the degree sequence $\{k_i\}$, connecting the ``stubs'' at random. The process to generate random clustered networks is quite similar. Let $s_i$ and $t_i$ be the number of single edges and the number of triangles attached to the node $i$, respectively. Given a network the sequence $\{s_i,t_i\}$ is possible to connect the ``stubs'' in order to generate single edges and also to connect nodes in order to obtain complete triangles. Hence, it is convenient to define the joint degree distribution $p_{s,t_\triangle}$ of the network, which is the fraction of vertices connected to $s$ single edges and $t_\triangle$ triangles. Therefore, the conventional degree of each node is given by $k = s + 2t_\triangle$, since each triangle contributes with two to the degree. Also, it is possible to relate the joint degree distribution $p_{st_\triangle}$ with the conventional degree distribution $p_k$ through
\begin{equation}
p_k = \sum_{s,t=0}^{\infty} p_{st_\triangle}\delta_{k,s+2t_\triangle},
\label{Eq:p_k}
\end{equation}
where $\delta_{i,j}$ is the Kronecker delta.

With the joint degree distribution $p_{s,t_\triangle}$ and the degree distribution $p_k$, we can calculate the clustering coefficient for random networks. The number of triangles in the network is given by $3N_{\triangle} = N \sum_{st}tp_{st}$ and the number of connected triples $N_{3} = N \sum_{k} \binom{k}{2}p_k$. Thus, using Eq.~\ref{Eq:clustering_coeff}, the clustering coefficient is
\begin{equation}
C = \frac{\sum_{st}tp_{st_\triangle}}{\sum_{k} \binom{k}{2}p_k}.
\label{Eq:clustering_coeff_2}
\end{equation}   
Note that the of factors $N$ cancel, letting $C>0$ in the limit of large networks, i.e., $N\rightarrow \infty$.

\section{Synchronization on clustered networks} 

The Kuramoto model consists of a set of $N$ oscillators coupled by the sine of their phase differences~\cite{strogatz2000kuramoto,Acebron05:RMP}. The state of each oscillator is characterized by its phase $\theta_i(t)$ $i=1,\ldots,N$. Considering a complex network where each node is a Kuramoto oscillator, the equations of motion are given by
\begin{equation}
\frac{d\theta_i(t)}{dt} = \omega_i + \lambda \sum_{i=1}^{N}A_{ij}\sin\left(\theta_j - \theta_i \right),\;i=1,\ldots,N,
\label{Eq:Kuramoto}
\end{equation}
where $\omega_i$ is the natural frequency of the node $i$ , $\lambda$ is the coupling strength and $A_{ij}$ are the elements of the adjacency matrix $\mathbf{A}$, where $A_{ij}=1$ if the nodes $i$ and $j$ are connected while $A_{ij}=0$, otherwise. The synchronization can be quantified through the order parameter
\begin{equation}
r e^{i\psi(t)} = \frac{1}{N} \sum_{j=1}^{N} e^{i\theta_j(t)},
\label{Eq:r}
\end{equation}
where $\psi(t)$ is the average phase of the system. The coherence parameter is bounded as $0\leq r \leq 1$, where $r=1$ represents the fully synchronized state and $r=0$ is the incoherent solution. In the fully connected graph ($A_{ij}=1\;\forall i,j$ and $i\neq j$), the order parameter $r$ as a function of $\lambda$ displays a second-order phase transition characterized by the critical coupling $\lambda_c = 2/(\pi g(\bar{\omega}))$~\cite{strogatz2000kuramoto,Acebron05:RMP}, where $g(\omega)$ is the distribution of the natural frequencies and $\bar{\omega}$ is the average frequency. In random networks, the critical coupling $\lambda_c$ of such phase transition is rescaled by the ratio $\left\langle k \right\rangle/\left\langle k^{2} \right\rangle$~\cite{Arenas08:PR,Ichinomiya04:PRE,lee2005synchronization,Restrepo05:PRE}, i.e., 
\begin{equation}
\lambda_c = \frac{2}{\pi g(\bar{\omega})} \frac{\left\langle k \right\rangle}{\left\langle k^{2} \right\rangle},
\label{Eq:critical_coupling_random_networks}
\end{equation}
where $\left\langle k^{n} \right\rangle$ is the $n$-th moment of the degree distribution $p_k$ of the network.

\subsection{Mean-field theory}

The critical coupling strength necessary for the onset of synchronization in Eq.~\ref{Eq:critical_coupling_random_networks} was firstly obtained by Ichinomiya through a mean-field analysis~\cite{Ichinomiya04:PRE} for the standard configuration model. The mean-field analysis has the advantage of allowing an analytical treatment. In order to extend the mean-field treatment to the configuration model for random clustered networks, we approximate the Eqs.~\ref{Eq:Kuramoto} into the following equation
\begin{equation}
\frac{d\theta_i(t)}{dt} = \omega_i + \lambda \sum_{k'}k_i P(k'|k_i)\sin\left(\theta_{k'} - \theta_i \right),
\label{Eq:Kuramoto_p_k}
\end{equation}
where $k_i$ is the degree of the node $i$ and $P(k'|k)$ is the probability that an edge emitted by a node with degree $k$ is connected to a node with $k'$. For uncorrelated networks $P(k'|k)$ is given by
\begin{equation}
P(k'|k) = \frac{k'p_{k'}}{\left\langle k \right\rangle}.
\label{Eq:p_k_lin_k}
\end{equation}
Substituting Eq.~\ref{Eq:p_k_lin_k} in Eq.~\ref{Eq:Kuramoto} we have:
\begin{equation}
\frac{d\theta_i(t)}{dt} = \omega_i + \frac{\lambda k_i}{\left\langle k \right\rangle} \sum_{k'} k'p_{k'}\sin\left(\theta_{k'} - \theta_i \right).
\label{Eq:Kuramoto_2}
\end{equation}
\begin{figure}[!t]
\centerline{\includegraphics[width=1\linewidth]{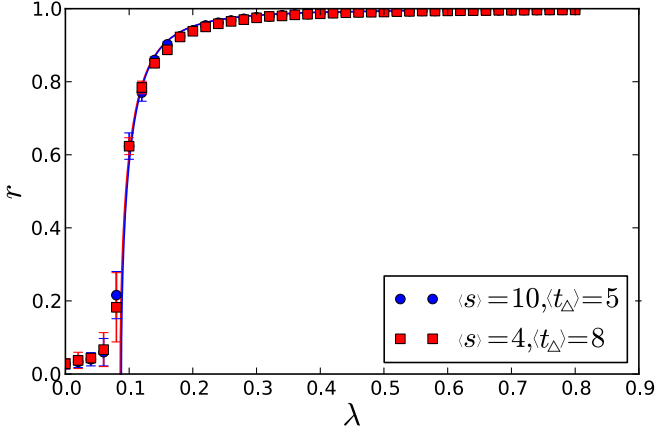}}
  \caption{Synchronization diagrams for networks with double Poisson joint distribution $p_{st_\triangle}$ (Eq.~\ref{Eq:pst_poisson}). The dots are obtained calculating the equations of motion (Eq.~\ref{Eq:Kuramoto}) until the system reaches the stationary state for each value of coupling $\lambda$. The order parameter $r$ is then calculated with Eq.~\ref{Eq:r}. Each point is an average over 10 network realizations. Solid lines correspond to the theoretical prediction from Eq.~\ref{Eq:r_implicit_equation}.}
  \label{fig1}
\end{figure}
For the configuration model of clustered random networks, $p_k$ is defined by Eq.~\ref{Eq:p_k}. Therefore, substituting in Eq.~\ref{Eq:Kuramoto_2} and noting that $k_i=s_i+2t_i$ we get 
\begin{equation}
\frac{d\theta_i(t)}{dt} = \omega_i + \frac{\lambda (s_i + 2t_i)}{\left\langle k \right\rangle} \sum_{s',t'_\triangle} (s'+2t'_\triangle)p_{s't'_\triangle}\sin\left(\theta_{k'} - \theta_i \right),
\label{Eq:Kuramoto_3}
\end{equation}
where $\left\langle k \right\rangle=\left\langle s \right\rangle +2\left\langle t_\triangle \right\rangle$. 

For an analytic treatment it is convenient to use the continuum limit of Eq.~\ref{Eq:Kuramoto_3}. For this purpose, let us define the density of the nodes with phase $\theta$ at time $t$, for a given $\omega$, with $s$ singles edges and $t_\triangle$ triangles, denoted by $\rho(s,t_\triangle,\omega;\theta,t)$. This density is  normalized as 
\begin{equation}
\int_0^{2\pi} \rho(s,t_\triangle,\omega;\theta,t) d\theta = 1.
\label{Eq:normalized}
\end{equation}
Therefore, Eq.~\ref{Eq:Kuramoto_3} in the continuum limit is given by
\begin{eqnarray}\nonumber
\frac{d\theta(t)}{dt} &=& \omega + \frac{\lambda (s + 2t_\triangle)}{\left\langle k \right\rangle} \int ds' \int dt'_\triangle \int d\theta' (s'+2t'_\triangle)\\
&& \times p_{s't'_\triangle}\sin\left(\theta' - \theta \right).
\label{Eq:Kuramoto_continuum}
\end{eqnarray}
The order parameter can also be redefined in order to account the connectivity pattern of a random network as
\begin{eqnarray}\nonumber
re^{i\psi(t)} & = & \frac{1}{\left\langle k\right\rangle }\int d\omega\int ds\int dt_{\triangle}\int d\theta(s+2t_{\triangle})p_{st_{\triangle}}\\
 &  & \times\rho(s,t_{\triangle},\omega;\theta,t)e^{i\theta}.
\label{Eq:r_II}
\end{eqnarray}
 
\begin{figure}[!t]
\centerline{\includegraphics[width=1\linewidth]{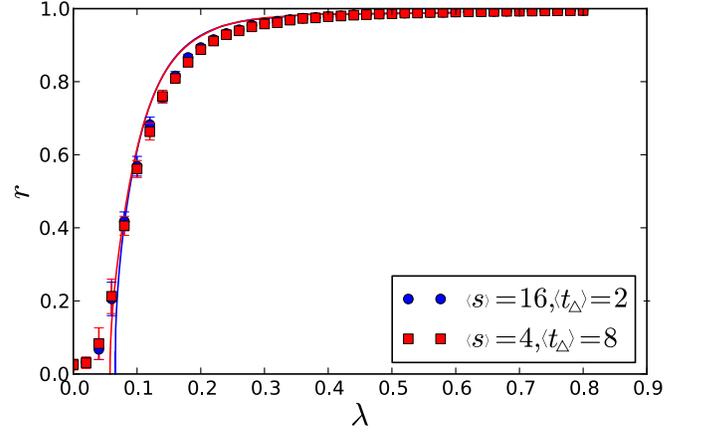}}
  \caption{Synchronization diagram calculated as in Fig.~\ref{fig1} for networks with double power-law joint distribution $p_{st_\triangle}$, with $\gamma_s=\gamma_{t}=3$ in  Eq.~\ref{Eq:pst_power_law}. Each point is an average over 10 network realizations. Solid lines correspond to the theoretical prediction from Eq.~\ref{Eq:r_implicit_equation}.}
  \label{fig2}
\end{figure} 
Considering Eq.~\ref{Eq:r_II}, it allows to rewrite Eq.~\ref{Eq:Kuramoto_3} in terms of the order parameter, resulting in
\begin{equation}
\frac{d\theta}{dt} = \omega + \lambda \left(s + 2t_\triangle \right)\sin(\psi - \theta).
\label{Eq:dtheta_dt_mean_field}
\end{equation}
The density $\rho(s,t_\triangle,\omega;\theta,t)$ will obey the following continuity equation
\begin{equation}
\frac{\partial \rho}{\partial t} + \frac{\partial}{\partial \theta}\{v_{\theta}\rho(s,t_\triangle,\omega;\theta,t) \}=0,
\label{Eq:cont_equation}
\end{equation}
which for the stationary states ($\partial\rho/\partial t=0$) has the solutions
\begin{equation}
\rho(s,t_\triangle,\omega;\theta)=\begin{cases}
\delta\left[\phi-\arcsin\left(\frac{\omega}{(s+2t_{\triangle})\lambda r}\right)\right] & \mbox{ if }\frac{\left|\omega \right|}{(s+2t_\triangle)}\leq \lambda r,\\
\frac{C(s,t_\triangle,\theta)}{\left|\omega-\lambda kr\sin\theta\right|} & \mbox{otherwise.}
\end{cases}
\label{eq:prob_solutions}
\end{equation}

These solutions correspond to those oscillators that are entrained by the mean field and those non-entrained, respectively. Thus, separating each contribution to the order parameter we yield
\begin{eqnarray}\nonumber
\left\langle k\right\rangle r & = & \int ds \int dt_\triangle d\theta \left[\int_{\frac{\left|\omega\right|}{(s+2t_\triangle)}\leq\lambda r}
 +  \int_{\frac{\left|\omega\right| }{(s+2t_\triangle)}>\lambda r}\right]\\
&\times &  p_{st_\triangle} (s+2t_\triangle)g(\omega)\rho(s,t_\triangle;\theta)e^{i\theta}.
\label{Eq:r_contributions}
\end{eqnarray}
The part of the non-entrained oscillator is given by 
\begin{eqnarray}\nonumber
\int d\theta g(\omega)e^{i\theta}\left[\int_{\lambda(s+2t_{\triangle})r}^{\infty}d\omega\frac{1}{(\omega-\lambda(s+2t_{\triangle})r\sin\theta)}\right.\\
+\left.\int_{-\infty}^{-\lambda(s+2t_{\triangle})r}d\omega\frac{1}{(-\omega+\lambda(s+2t_{\triangle})r\sin\theta)}\right]=0,
\label{Eq:non-entrained}
\end{eqnarray}
since the integral over $\theta$ is 0.  Thus the only contribution for coherence the parameter $r$ is due to the synchronous oscillators, which is accounted in Eq.~\ref{Eq:r_contributions}:
\begin{eqnarray}\nonumber
\left\langle k\right\rangle r &=&\int ds\int dt_{\triangle}\int d\omega\int d\theta(s+2t_{\triangle})g(\omega)\\
&& \times p_{st_{\triangle}}\exp\left[i\arcsin\left(\frac{\omega}{(s+2t_{\triangle})\lambda r}\right)\right].
\label{Eq:r_sync_contr}
\end{eqnarray}
From the real part we get:
\begin{eqnarray}\nonumber
\left\langle k\right\rangle r &=&\int ds\int dt_{\triangle}\int d\omega\int d\theta(s+2t_{\triangle})g(\omega)\\
&& \times p_{st_{\triangle}}\sqrt{1 - \left(\frac{\omega}{(s+2t_{\triangle})\lambda r}\right)^{2} }.
\label{Eq:r_real_part}
\end{eqnarray}
Considering the following change of variable $\omega' = \omega/(s+2t_\triangle)\lambda r$ and considering $g(\omega)=(\sqrt{2\pi})^{-1}e^{-\omega^{2}/2}$, we obtain the following implicit equation for the coherence parameter $r$ 
\begin{widetext}
\begin{equation}
\lambda=\sqrt{\frac{\pi}{8}}\left\langle k\right\rangle \left\{ \int\int(s+2t_{\triangle})^{2}p_{st_{\triangle}}e^{-\lambda^{2}(s+2t_{\triangle})^{2}r^{2}/4}\left[I_{0}\left(\frac{\lambda^{2}(s+2t_{\triangle})^{2}r^{2}}{4}\right)+I_{1}\left(\frac{\lambda^{2}(s+2t_{\triangle})^{2}r^{2}}{4}\right)\right]dsdt_{\triangle}\right\} ^{-1},
\label{Eq:r_implicit_equation}
\end{equation}
\end{widetext}
where $I_0$ and $I_1$ are the modified Bessel functions of first kind. Thus, tending $r^+ \rightarrow 0$, we obtain the critical coupling $\lambda_c$ for the onset of synchronization
\begin{equation}
\lambda_c = \sqrt{\frac{8}{\pi}}\left\langle k \right\rangle \left\{\int \int (s+2t_\triangle)^2 p_{st_\triangle} dsdt_\triangle \right\}^{-1}.
\label{Eq:critical_coupling_cluster}
\end{equation}
The mean field result obtained for the critical coupling $\lambda_c$ for clustered netwokrs, Eq.~\ref{Eq:critical_coupling_cluster}, is similar to the one for non-clustered networks (Eq.~\ref{Eq:critical_coupling_random_networks}), i.e., in both cases the critical coupling is proportional to the ratio of the moments of the degree distribution. Note that in the absence of triangles $(t_\triangle=0)$ we recover the result $\lambda_c = \sqrt{8/\pi} \left\langle k \right\rangle/ \left\langle k^2 \right\rangle$, where the degrees are just due to single edges, $k=s$. 

\begin{figure}[!tpb]
\centerline{\includegraphics[width=1\linewidth]{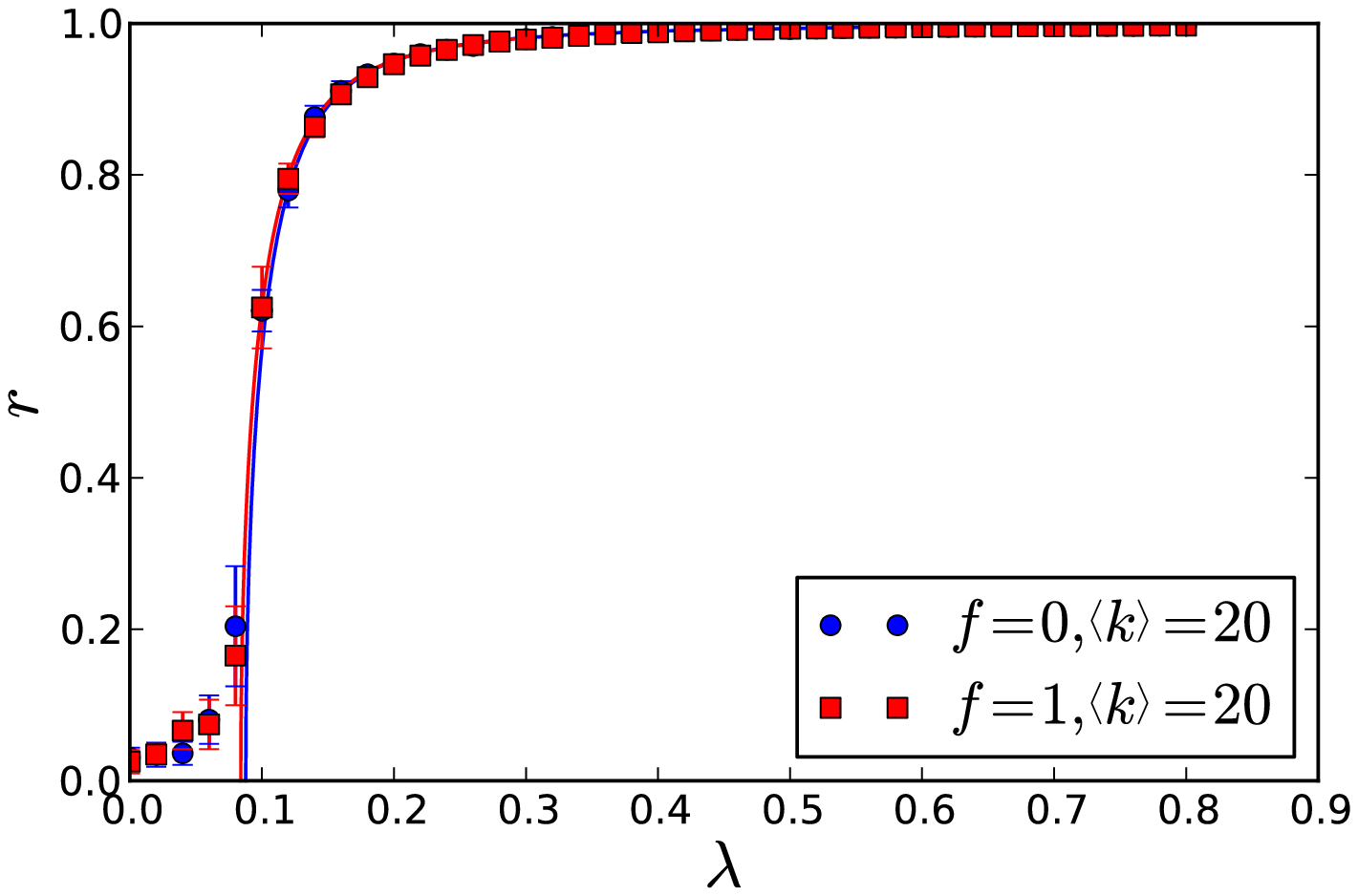}}
  \caption{Synchronization diagrams calculated as in the other figures for random networks with degree distribution $p_k=e^{-\left\langle k \right\rangle}\left\langle k \right\rangle ^{k}/k! $ and joint degree distribution generated by  Eq.~\ref{Eq:pst_f}. Each point is an average over 10 network realizations. Solid lines correspond to the theoretical prediction from Eq.~\ref{Eq:r_implicit_equation}.}
  \label{fig3}
\end{figure}
\section{Numerical simulations}

In this section we give some numerical simulations of clustered and non-clustered networks and will compare them with the theoretical result of Eq.~\ref{Eq:r_implicit_equation}. All simulations consider networks which are constructed through the configuration model presented in Sec.II and the distributions of single edges and triangles independently. However, it is also possible consider correlated distributions as well.

\begin{figure}[!tpb]
\centerline{\includegraphics[width=1\linewidth]{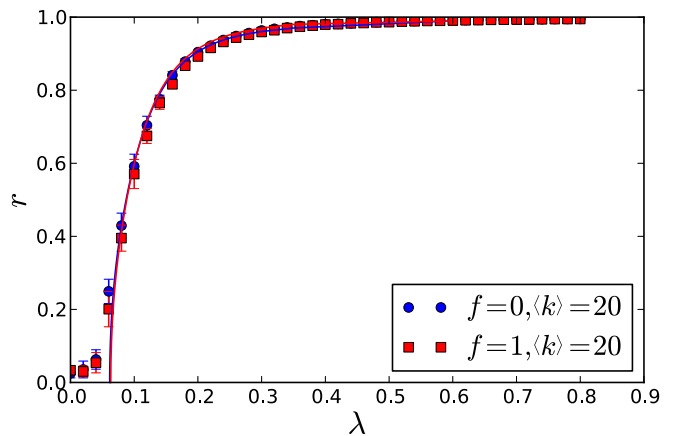}}
  \caption{Synchronization diagrams calculated as in the other figures for random networks with degree distribution $p_k \propto k^{-\gamma} $, with $\gamma=3$ and joint degree distribution generated by  Eq.~\ref{Eq:pst_f}. Each point is an average over 10 network realizations. Solid lines correspond to the theoretical prediction from Eq.~\ref{Eq:r_implicit_equation}.}
  \label{fig4}
\end{figure} 
Let us study first networks with the following joint distribution of single edges and triangles
\begin{equation}
p_{st_\triangle} = e^{-\left\langle s \right\rangle}\frac{\left\langle s \right\rangle^{s}}{s!} e^{-\left\langle t_\triangle \right\rangle}\frac{\left\langle t_\triangle \right\rangle^{t_\triangle}}{t_\triangle!}.
\label{Eq:pst_poisson}
\end{equation}
In order to analyse systematically the dependence of the order parameter on the presence of triangles in the network, we kept the average degree $\left\langle k \right\rangle=\left\langle s \right\rangle + 2\left\langle t_\triangle \right\rangle$ fixed and varied the $\left\langle s \right\rangle$ and $\left\langle t_\triangle \right\rangle$, calculating the order parameter $r$ as a function of the critical coupling $\lambda$. Fig.~\ref{fig1} shows the synchronization diagram for networks with double Poisson degree distributions (Eq.~\ref{Eq:pst_poisson}) with average degree $\left\langle k \right\rangle = 20$. It is interesting to note that networks with higher values of $\left\langle t_\triangle \right\rangle$ have the same critical coupling for the onset of synchronization.

We have also considered networks with joint distribution consisting of a double power-law distribution 
\begin{equation}
p_{st_\triangle} \propto s^{-\gamma_s} t_\triangle ^{-\gamma_t}, 
\label{Eq:pst_power_law}
\end{equation}
where $\gamma_s = \gamma_t = \gamma$ for the sake of simplicity. Fig.~\ref{fig2} shows order parameter $r$ as a function of $\lambda$ considering $\gamma=3$. As we can see, the same behavior is observed as in Fig.~\ref{fig1}, the presence of clustering in the network does not affect the network synchronization. The non-zero values of the order parameter $r$ for small values of the coupling $\lambda$ in Fig.~\ref{fig1} and~\ref{fig2} are due to finite-size effects~\cite{Ichinomiya04:PRE,lee2005synchronization,Restrepo05:PRE}.

It is also possible to construct the joint distribution $p_{st_\triangle}$ from a given degree distribution $p_k$ through the relation~\cite{hackett2011cascades,gleeson2010clustering}
\begin{equation}
p_{st_\triangle} = p_k \delta_{k,s+2t_\triangle}\left[(1-f)\delta_{t,0}+f\delta_{t,\left\lfloor (s+2t_{\triangle})/2\right\rfloor }\right]
\label{Eq:pst_f}
\end{equation}
where $0\leq f \leq 1 $ and $\left\lfloor \cdot \right\rfloor$ is the floor function. Through Eq.~\ref{Eq:pst_f} we can construct $p_{st_\triangle}$ keeping the degree distribution $p_k$ fixed with $f$ being the fraction of nodes in the network attached to the maximum possible number of triangles $t = \left\lfloor (s+2t_\triangle)/2 \right\rfloor$ and $(1-f)$ the fraction of nodes which are attached to single edges only. Substituting Eq.~\ref{Eq:pst_f} into Eq.~\ref{Eq:clustering_coeff_2} we obtained~\cite{hackett2011cascades}

\begin{equation}
C=f\frac{\sum_{k}(p_{2k}+p_{2k+1})}{\sum_{k}\binom{k}{2}p_{k}}.
\label{Eq:clustering_coeff_3}
\end{equation}
Eq.~\ref{Eq:clustering_coeff_3} establishes a linear relationship between $C$ and $f$, i.e., with $f=0$ we construct a network with the minimum value for the cluster coefficient and $f=1$ a network with the maximum value of $C$ for a given $p_k$, allowing to study the extreme cases of the topology.  Fig.~\ref{fig3} and Fig.~\ref{fig4} show the synchronization diagrams for networks with $p_k=e^{-\left\langle k \right\rangle}\left\langle k \right\rangle ^{k}/k! $ and $p_k \propto k^{-\gamma} $, respectively.
Again, we observe a good agreement with the theoretical curve. Therefore, the clustering coefficient has no effect on the coherence parameter evolution $r(\lambda)$, comparing the curves with $f=0$ and $f=1$. Also, finite-size effects are observed for small values of $\lambda$.

In summary, we have shown that the presence of cycles of order three does not play an important role in network synchronization of Kuramoto oscillators. In fact, the theoretical results for non-clustered networks are highly accurate on describing the behavior of the order parameter $r$ for clustered networks, even when the cluster coefficient $C$ has the maximum accessible value for a given network. The results presented here are in agreement with previous findings~\cite{melnik2011unreasonable}, where it was found that the presence of triangles in the network topology does not influence the performance of other dynamical processes, such as bond percolation, $k$-core size percolations and epidemic spreading.

F. A. Rodrigues would like to acknowledge CNPq (305940/2010-4) and FAPESP (2010/19440-2) for the financial support given to this research. T. Peron would like to acknowledge FAPESP and J. Kurths would like to acknowledge IRTG for the sponsorship provided.

\bibliographystyle{apsrev}
\bibliography{paper}

\end{document}